# LEARNING CONTEXT FOR TEXT CATEGORIZATION

Y.V. Haribhakta[1] and Dr. Parag Kulkarni[2]

[1]Department of Computer Engineering & I.T. , College of Engineering, Pune, Maharashtra, India
ybl.comp@coep.ac.in
[2] EkLat Labs, Pune, Maharashtra , India
paragindia@gmail.com

## ABSTRACT

*This paper describes our work which is based on discovering context for text document categorization. The document categorization approach is derived from a combination of a learning paradigm known as relation extraction and an technique known as context discovery. We demonstrate the effectiveness of our categorization approach using reuters 21578 dataset and synthetic real world data from sports domain. Our experimental results indicate that the learned context greatly improves the categorization performance as compared to traditional categorization approaches.*

## 1. INTRODUCTION

Text Categorization has become an active research topic in the area of machine learning. The task of text categorization is to classify a document under a predefined category. A document refers to piece of text. Categories may be derived from a sparse classification scheme or from a large collection of very specific text documents. Categories may be represented numerically or using single word or phrase or words with senses, etc. In traditional approach, categorization task was carried out manually using domain experts. Each incoming text documents was read and comprehended by the experts and assigned to one or more number of categories chosen from the set of predefined categories. It is inevitable that enormous human efforts was required.

  A perfect true way to handle this problem is to learn an automated categorization scheme from training examples. Once the categorization scheme is learned , it can be used to classify future uncategorized documents. There are several issues involved in this which are normally found in many machine learning problem. The scheme should be able to handle multiple category assignment for a document as a document may be assigned to more than one category. There is a large research community addressing automatic text categorization. For instance, CogCate[1] introduces an innovative content-oriented text categorization which exploits a human cognitive procedure in categorizing texts. It applies lexical/ semantical analysis in addition to traditional statistical analysis at word which ensures the accuracy of categorization. [2] proposes a fuzzy ranking analysis paradigm together with a novel relevance measure, discriminating power measure (DPM), to effectively re-duce the input dimensionality from tens of thousands to a few hundred with zero rejection rate and small decrease in accuracy . Automatic Web Page Categorization by Link and Context Analysis [3] paper describes the novel technique of categorization by context, which instead extracts useful information for classifying a document from the context where a URL referring to it appears. Categorization by context exploits an essential aspect of a hypertext environment like the Web, structure of the document and the link topology . [4] proposes a more robust algorithm for keyword extraction to induce concepts from training examples, which is based on enumeration of all possible keywords combinations .WordSieve [7,9] an unsupervised term extraction algorithm suggest that it is a promising approach for extracting key terms for indexing documents according to the contexts in which they are used and for differentiating a users' different task contexts. It also suggests that is guided by the hypothesis that the relevant features of a document depend not only on what makes it different from every other document (which is captured by TFIDF), but what makes it similar to the documents with which it was accessed. In other words, the context of a document is not reflected only in its content, but in the other documents with which it was accessed. [16,20,21] discusses techniques of efficient association rule mining for extracting relations from text. In the early work of Lewis[22], a probabilistic model makes use of bayesian independent classifiers for categorization. The model showed the effect of feature selection and clustering on the automatic categorization of newswire articles. Yang[19] developed a technique known as expert network. This network links the terms in a document with its categories and each link has a weight assigned to it. Other methods such as context sensitive learning [14] , linear classifier [15], learning by combining classifier [17] and decision tree [18] have also been proposed. These approaches typically construct a classifier for each category and the categorization process becomes a binary decision problem for the particular category. In contrast, our approach learns all the categories for a document at one time. We experimented our classifier on reuters 21578 dataset and synthetic real world data from sports and politics domains.

This paper is organized in two parts. Part I focuses on category training model and part II focuses on learning context for categorization. For part I , the Section 2 describes the category training approach followed by the model for score calculation of features returned by the association rule mining algorithms. The Part II describes the model for learning context. This is followed by experimentation and results on two document collection , namely, reuters 21578 and sport test collections. Finally , section 5 and 6 provides the conclusion and future research work.

## 2. PART I : CATEGORY TRAINING MODEL

### 2.1. An outline of the category training model

The category training model provides an algorithm to train the collection of text documents in their respective categories. The algorithm consists of two processes, namely, relation extraction and score calculation for extracted features of relations. Relation Extraction is essentially a classification problem. The task of relation extraction aims to establish relations between the classified entities. For classifications of the documents, we are extracting relations from the document. To extract the relevant relations from the documents we are using the concept of association rule mining. Association rule mining, one of the most important and well researched techniques of data mining, was first introduced in [1]. It aims to extract interesting correlations, frequent patterns, associations among set of items in the transaction databases or other repositories. Association rule mining is to find out association rules that satisfy the predefined minimum support and confidence from a collection. The second process of score calculation for extracted features of relations assigns score to every feature with respect to the context in which it occurs .
The objective of the category training model is to select appropriate context(pre-defined) for the input document. A collection of pre-defined context documents are used for training. Each document ,which contains a free-text portion, is trained using the algorithm given below for every defined context . Our approach learns all contexts for a document at one time. A document in the training collection is considered as an instance represented by < D;C> where D represents the free text and C represents the contexts of the document.

The vector space technique is adopted as the central representation framework for the model. Thus, D is vector of terms which are the features: D = { $f_1$, $f_2$, ……………, $f_n$} where n is the total number of unique features in the collections free-text domain and $f_i$ is the weight reflecting the relative importance of the feature i for characterizing the document. Typically, the features are the highly relevant terms of the text obtained using association rule mining. Similarly, C is a vector representing the contexts assigned to the document i.e., C = { $c_1$, $c_2$, ……., $c_m$} where $c_i$ is the weight of the context i and m is the total number of unique contexts. A number of weighting schemes can be used for the vectors D and C. For instance, we can use the product of term frequency and inverse document frequency as the weights for the features extracted in D. Term frequency is the frequency count of the feature in the document. Inverse document frequency is related to the rarity of the feature in the document collection. The score calculation model, given below, assigns weight to every feature of the set D.

## 2.2. Relation Extraction

### Formal Problem Description

Let I = { $x_1, x_2, ………..,x_n$} be a set of distinct terms called items. A set X ⊆ Y with k = |X| is called a k-itemset or simply an itemset. Let a database D be a multiset of subsets of I. Each T ⊆ D is called a transaction. We say that a transaction T ⊆ D supports an itemset X U I if X ⊆ T holds. An association rule is an expression X ⇒ Y , when X and Y are the itemsets and X ∩ Y = Φ holds. The fraction of transactions T supporting an itemset X with respect to database D is called the support of X, supp(X) = {| T ∈ D | X ⊆ T }| / | D| . The support of a rule X ⇒ Y is defined as supp(X ⇒ Y) = supp ( X ∪ Y) . The confidence of this rule is defined as conf(X ⇒ Y) = supp(X ∪ Y) /supp(X). [7].

Our approach for relation extraction uses different association rule mining algorithms[6][7][8][9] , to extract the interesting relations for the text document. The input text document is preprocessed to a set of transactions. Each sentence of the text document is interpreted as an transaction. The itemsets from the relations extracted are the features considered for designing context feature matrix required for part II of the algorithm. The algorithms used to extract the relations are Apriori, MSapriori, RSApriori and Diffset. The figure 1 shows the experimental results of all these algorithms. The minimum support considered is 5% of the total number of transactions. For RSApriori, minimum support for rare itemsets is 3% and relative support is 0.6.

Figure 1: Performance analysis of relation extraction algorithms

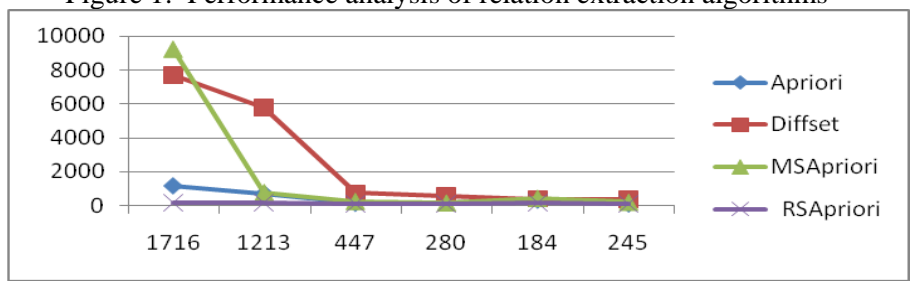

In the figure 1, the X-axis has number of transactions and the Y-axis has time. It can be observed that for less number of transactions all algorithms work similar but as the number of transactions increases the performance of each algorithm differs. Hundreds of documents from the web are collected and results are calculated after rigorous testing of all these algorithms. With MSApriori algorithm, as the number of transactions increases above 500, there is sudden rise in the graph due to calculation of MIS value for each set of itemset whereas in the Diffset algorithm the sudden rise in the graph is due to the set theory calculations with the increase of number of transactions. Diffset is a vertical mining algorithm and resulting valid n-itemsets are with higher value of n. The result shows Apriori performs better in time compared to Diffset and MSApriori, but it has the rare itemset dilemma. RSApriori is better than MSApriori in time complexity. Both RSApriori and MSApriori handles the rare itemset dilemma which is ignored in Apriori. RSApriori handles rare itemset dilemma with better approach than MSApriori.

### 2.3. Weight Calculation of extracted features

The extracted features of relation extraction are the items within the itemsets for the relations extracted. A particular feature $f_i$ is relevant to a document dj if its occurrence is more in it. The term relevance is also measured looking at the occurrence of that feature in different documents for a single cluster. To quantify the relevance factor, calculate the frequency of each distinct feature in different documents for a single cluster. This model provides an algorithm to assign score to every feature for a collection of similar type documents. The algorithm for score calculation is summarized as below :

---

Input : Text Document
Output : Features with score for every cluster ( where cluster defines a context)

From the collection of training instances
1. Group similar type of documents in single cluster.
2. Call association_rule_mining algorithms to extract relations and eventually features.
3. Calculate frequency of each distinct extracted feature within every cluster.
4. Group identical frequency features in single cluster .
5. Arrange all clusters according to frequency.
6. Assign indices to every cluster, with index 0 to highest frequency cluster and successively assigning the indices to others.
7. Weight of a feature $f_i$ in a document $d_j$ for a cluster $C_k$ is calculated as :

$$W_{(f_i, d_i)} = 1 - ( \gamma / n ) \qquad\qquad -- (1)$$

where,
$W_{(f_i, d_j)}$ : Weight of the feature $f_i$ in a document $d_j$
$\gamma$ : index of the cluster to which the feature $f_i$ belongs.
n : total number of cluster
$C_k$ : $k^{th}$ cluster of similar documents ($d_j, d_{j+1}, --, d_k$)

8. Initially, Score Calculation for the feature $f_i$ if it belongs to document $d_j$ for a single cluster $C_k$ is calculated as :

$$Score (f_i, d_j) = W_{(f_i, d_i)} * D_j \qquad\qquad -- (2)$$

$$= W_{(f_i, d_i)}$$

( If $d_j$ is the first document of the cluster i.e., $D_j = 1$ )

9. Iteratively, if $f_i$ belongs to the next document $d_{j+1}$ in the cluster $C_k$, the score is calculated as:

$$Score (f_i, d_{j+1}) = Score (f_i, d_j) * D_j + Score(f_i, d_{j+1}) / D_j + 1 \qquad\qquad -- (3)$$

where, $D_j$ is the document frequency before considering the document $d_{j+1}$ for calculating the score of the feature $f_i$.

The steps 1 through 9 are repeated for every cluster (which defines the context) for score calculation of extracted features. So, finally we have a trained dataset for each context which consists of features with their scores or weight.

## 3. PART II : LEARNING CONTEXT FOR CATEGORIZATION

Context is the theme of the document. It is a set of facts (topics or occurrences) grouped on the basis of their relations to a common set of topics, associations, or scopes describing a given goal. In the same context all documents share a common association type. The objective of this model is to discover or learn the appropriate context for an input document. The context discovery process involves two processes, namely, constructing context feature matrix and extracting appropriate context.

Let Q denotes an incoming document which needs to be learned for context. Since free-text portion of the document is available, we represent Q by a vector of relevant unique features :

$$Q = \{ q_1, q_2, \ldots\ldots\ldots\ldots, q_n \}$$

where $q_i$ is the relevant feature in Q obtained using the association rule mining algorithm. Then, the context feature matrix is constructed by considering all the distinct relevant features of the vector Q. Let : CFM= C x Q be a context feature matrix where C is the set of predefined contexts and Q is the set of extracted features. Each entry of the matrix will have a value in the range 0-1. The value at the position CFM[ i, j ] indicates importance of the $j^{th}$ feature in the $i^{th}$ context. A value of 0 indicates the feature is irrelevant for the context. A value nearer to 1 or 1 indicates the feature is highly relevant for the context, otherwise the feature is average relevant. The value for each entry of the CFM is the score value obtained from the trained dataset for all contexts. The score value is nothing but the weight signifying the importance of that feature for a context.

Finally, the context for the input document is extracted by looking at the score values of the features across all contexts. For each feature, summation of the score values across all context is performed. That context which has the highest score is set as context of the input document.

The Algorithm for context discovery is summarized as below :

```
Input : Text Document
Output : Context of the document

For each input text document
 {
  call relation_extraction_algorithm { to extract terms which are relevant features }
  construct context_feature_matrix
 }
 for each context
Context Score = ∑ score (relevant features)
            q_i ЄQ
Context ( Input document ) = max (Context Score) of all contexts
```

## 4. EXPERIMENTATION & RESULTS

The context discovery model has been trained and tested on the standard benchmark dataset reuters 21578 [17] and real world data from the sports domain. For experimentation, only five sub categories of the Topics category from reuters 21578 are considered, viz, acq, interest, ship, trade and earn. The category training model designs a weighted feature template for each sub category. These sub categories are selected as context for the Topics category. For the sports domains, 14 sport games are considered for training, viz, Archery, Badminton, Baseball,

Basketball, Chess, Cricket, Golf, Rugby, Soccer, Squash, Table Tennis, Tennis, Volley, Water Polo . Nearly 2000 sports documents collected from google and yahoo newsgroup are considered for training. The 14 games specified above of the sports domain are selected as context. Most of the games have features that are unique for that game whereas have features that are common across some or all of the games. This property plays role in the learning of context for the document. Though the features are common across some or all of the games , depending upon the importance of those features for a particular game, they will be weighted accordingly. The benchmark testing dataset of reuters 21578 across the five sub categories of the Topics category (acq, interest, ship, trade and earn) are considered. Also, testing data from the sports domain is considered. Nearly 2000 sports document collected from google and yahoo newsgroup are considered for experimentation.

The testing results are compared with some of the benchmarking algorithms like naïve-bayesian, k-NN, SVM, and Rochhio( ) to show the performance in terms of precision, recall and f-measure.

Precision (P) = true positive($t_P$) / true positive ($t_P$) + false positive($f_P$)

Recall (R) = true positive($t_P$) / true positive ($t_P$) + false negative ($f_N$)

where,

$t_P$ is the number of documents correctly labeled as belonging to the positive class

$f_P$ is the number of documents incorrectly labeled as belonging to the positive class

$f_N$ is the number of documents which were not labeled as belonging to the positive class but should have been

F-measure is the harmonic mean of precision and recall,

F-measure = 2 * [(precision * recall )/(precision +recall)]

The figure 2 shows the F-measure performance calculated for collection of 280 articles of sports domain (randomly selected from the web) whereas figure 3 shows the F-measure performance of reuters 21578 across 5 topics category . We observe that the learning accuracy achieved is

| Table 1 : F-Measure Performance for Sports Domain | | | |
|---|---|---|---|
| | Precision | Recall | F-Measure |
| Archery | 0.85 | 1 | 0.92 |
| Badminton | 1 | 0.63 | 0.77 |
| Baseball | 0.8 | 0.57 | 0.67 |
| Basketball | 0.89 | 0.35 | 0.5 |
| Chess | 0.9 | 1 | 0.95 |
| cricket | 0.87 | 0.67 | 0.75 |
| Golf | 0.92 | 0.86 | 0.89 |
| Rugby Union | 0.78 | 0.82 | 0.8 |
| Soccer | 0.44 | 0.3 | 0.36 |
| Squash | 0.74 | 0.68 | 0.71 |
| tennis | 0.69 | 0.45 | 0.52 |
| table tennis | 0.4 | 0.82 | 0.54 |
| voleyball | 0.53 | 0.35 | 0.42 |
| Water Polo | 0.45 | 0.86 | 0.59 |

| Table 2 : F-Measure Performance for Topics Domain of Reuters 21578 | | | |
|---|---|---|---|
| | Precision | Recall | F-Measure |
| acq | 0.7795 | 0.7795 | 0.7795 |
| interest | 0.625 | 0.5 | 0.5556 |
| ship | 0.8462 | 0.4889 | 0.6197 |
| trade | 0.87 | 0.7429 | 0.8000 |
| earn | 0.68 | 0.58 | 0.6239 |

about 89-95% in sports such as chess, golf, and archery. Our learner classifies the document according to the unique features used in that particular sport, like features archery, golf are very rare and are uniquely define for archery and golf sports only. For instance, the features bow and arrow define archery game only, similarly features such as king, queen, castling define the sport chess. But there are some features which are very commonly used in various sports which confuses the learner like the features ball, score, move etc., which are very frequently used in games like cricket, water polo, soccer, basket ball,etc.

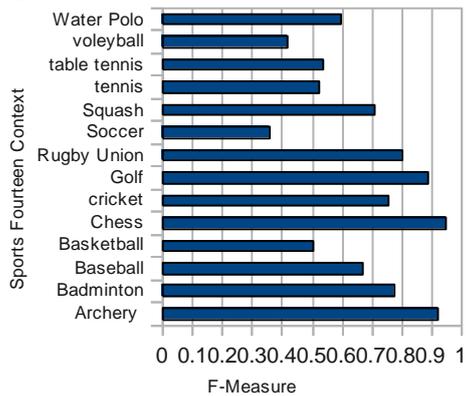

Fig.2: F-Measure Performance for Sports Domain

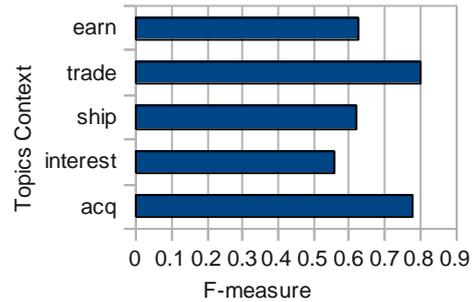

Fig.3: F-Measure Performance of Reuters 21578 For Topics Domain

The learner may get confused to set the context of the sport looking at these features. This problem is resolved by assigning weights to the common features. All these common features have different weights in different games. So, the category training model has assigned weights to these common features by looking at the importance of these features across all the game. But still soccer, basketball, badminton are the sports having many features similar and equally used. Hence the results acquired for these sports are 65-70% on an average. Other than these sports , other sports have given more than 78% correct results.

## 5. CONCLUSION

The current research has considered a supervised approach for determining context using association rule mining approach. The templates designed for determining context consists of collection of terms relevant for the context. Every term has associated with it a value which signifies the importance or weightage of that term in that context. For the features extracted for a document, weight is calculated which plays major role in determining the correct context. The future work can be , instead of considering single terms as features , a keyterm consisting of two or three terms should be considered for designing templates which might help in improving the accuracy of categorization. The parameters in the weighting model can be experimented with different values for much better performance .Currently sports and politics dataset are considered for experimentation. More experimentation should be done on different datasets for fine tuning the parameters..

## REFERENCES

[1] N. H. Yi Guo, Zhiqing Shao, "Automatic text categorization based on Content Analysis with Cognitive Situation Model," in Information Sciences. Science Direct, 2010, pp. 613–531.
[2] C.-M. Chen, "Two novel feature selection approaches for web page classification," in Expert Systems with Applications, Science Direct, 2009, pp. 260–273.

[3] F. S. Giuseppe Attardi, Antonio Gull, "Automatic web page categorization by link and context analysis," 2000.
[4] Y.-P. P. C. Jiyuan An, "Keyword extraction for text categorization." IEEE Computer Society, 2005, pp. 556–561.
[5] B. F. Andrej Bratko, "Exploiting structural information for SemiStructured document categorization," in Information Processing and Management. ACM, 2005.
[6] L. W. Y.-F. H. Xiao-Yun Chen, Yi Chen, "Text Categorization based on frequent patterns with term frequency." IEEE Computer Society, 2004.
[7] Travis L. Bauer ,David B. Leake , "Detecting Context-Differentiating Terms Using Competitive Learning ",SIGIR October 2, 2003.
[8] Mohmmed Zaki. Fast Vertical Mining using Diffset. SIGKDD'03, August 2003. Washington, DC, USA Copyright 2003 ACM.
[9] D. B. L. Travis Bauer, "Wordsieve : A method for real-time Context Extraction." IEEE Computer Society.
[10] Fabrizio Sebastiani. Machine Learning in Automated Text Categorization. ACM Computing surveys ,vol.34, No.1,March 2002, pp.1-47.
[11] V. H. Jihoon Yang, "Feature subset selection using a genetic algorithm," in ACM Computing Classification System Categories. ACM, 2000.
[12] J.D. Holt, S.M. Chung. Efficient Mining of Association Rules in Text Databases, CIKM'99, Kansas City, USA,pp.234-242(Nov 1999).
[13] J.S. Park, M.S. Chen and P.S.Yu. Using a Hash-based Method with Transaction trimming for Mining Association rules. IEEE Transactions on Knowledge and Data Engineering.Vol9, No.5,Sept/Oct, 1997.
[14] W.W. Cohen and Y. Singer. Context – Sensitive Learning Methods for Text Categorization. Proc 19th Int'l ACM SIGIR Conf. Research and Development in Information Retreival , pp.307-315, 1996.
[15] D.D. Lewis, R.E. Schapire, J.P. Callan and R. Papka. Training Algorithms for Linear text Classifiers. Proc. 19th Int'l ACM SIGIR Conf. Research and Development in Information Retrieval , pp. 298-306, 1996.
[16] R.Agrawal, H. Mannila, R. Srikant, H. Toivonen, and A. I, Verkamo. Fast discovery of association rules. In Advances in Knowledge Discovery and Data Mining, pages 307–328, 1996.
[17] L.S. Larkey and W.B. Croft. Combining Classifiers in Text Categorization. Proc. 19th Int'l ACM SIGIR Conf. Research and Development in Information Retrieval, pp.289-297, 1996.
[18] C. Apte, F. damerau, and S.M. Weiss. Automated Learning of Decision Rules for Text Categorization. ACM Transaction. Information Systems, Vol 12, no. 3, pp 233-251, 1994.
[19] Y. Yang. Expert Network: Effective and Efficient learning from Human Decisions in text categorization and Retreival. Proc. 17th Int'l ACM SIGIR Conf. Research and Development in Information retreival, pp.13-22, 1994.
[20] Agrawal R. , Srikanth R. Fast Algorithms for mining association rules VLDB, 1994.
[21] R.Agrawal, T. Imielinski, A. Swami. Mining Associations between Sets of Items in Massive Databases. Proceedings ACM SIGMOD 1993, pp. 207-216.
[22] D.D. Lewis. Feature Selection and Feature Extraction for Text Categorization. Proc. Speech and Natural Language Workshop, pp 212-217, 1992.
[23] R. Uday Kiran and P. Krishna Reddy . An Improved Multiple Minimum Support Based Approach to Mine Rare Association Rules.
[24] S.Ayse Ozel and H.Altay. An algorithm for Mining Association Rules using perfect hashing and database pruning. Bilkent University, Department of Computer Engineering, Ankara, Turkey.
[25] www.daviddlewis.com/resources/testcollections/reuters21578.

**Authors**

1)  **Yashodhara V. Haribhakta**
Working as Assistant Professor with the Department of Computer Engg. & I.T. , College of Engineering, Pune. Pursuing research in the area of Machine Learning. Areas of interest are Machine Learning, Text Mining and Natural Language Processing.

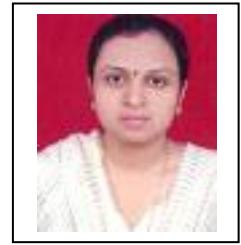

2)  **Dr. Parag Kulkarni**
Chief Scientist and Founder of EkLat Labs Pune Received Ph.D. degree in Computer Science and Engineering from IIT, Khargpur. Conferred Higher Doctorate - Doctor of Sciences (D.Sc.) for professional contribution and research in empowering businesses with Machine Learning, Knowledge Management and Systemic Management. Areas of interest are Machine Learning, Knowledge Management, and IT strategies.

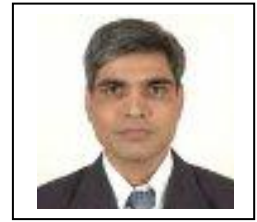